\documentclass[%
 reprint,
amsmath,amssymb,
aps,
]{revtex4-2}

\usepackage{dcolumn}
\usepackage{bm}

\usepackage{graphicx}
\usepackage{subfigure}
\usepackage{graphicx,hyperref}
\expandafter\let\csname equation*\endcsname\relax
\expandafter\let\csname endequation*\endcsname\relax
\usepackage{amsmath}
\usepackage{amssymb}
\usepackage{multirow, makecell, comment} 
\newcommand{\fla}[1]{\begin{flalign}#1\end{flalign}}
\usepackage{braket}
\usepackage{lineno}
\usepackage{lipsum}
\usepackage{xcolor}
\usepackage{soul}
\usepackage{array}

\begin{document}

\preprint{APS/123-QED}

\title{Deterministic synthesis and processing of frequency-bin qubits in a macroscopically coherent quantum memory
}

\author{  S.A. Moiseev}
\email{samoi@yandex.ru}

\affiliation{Kazan Quantum Center, Kazan National Research Technical University n.a. A.N. Tupolev-KAI, 10 K. Marx St., 420111, Kazan, Russia}

\date{\today}

\begin{abstract}

Quantum information processing requires efficient storage and manipulation of photonic states. In this work, we harness pre-created macroscopic spin coherence as an additional degree of freedom to control a cavity-assisted quantum memory. We derive equations of motion for such a memory that predict efficient emission and absorption of two-color fields under a generalized nonlocal spectral impedance-matching condition, which reduces to the well-known condition in the single-mode case. These equations reveal a new operating mechanism of cavity-assisted quantum memory wherein the controlled spin coherence couples the emitted spectral modes. As a result, our memory deterministically synthesizes arbitrary frequency-bin photonic qubits using simple radio-frequency rotations, while the temporal reversibility of the protocol provides perfect transformation of the qubit states. Our findings establish a unified platform bridging quantum storage with elementary signal processing for quantum networks based on spectral encoding of photonic qubits

\end{abstract}

                              
\maketitle

\section{Introduction}

The creation of optical quantum memory (QM) is aimed at solving a wide range of practical problems in the field of optical quantum information
\cite{Lukin2003,Lvovsky-NatPhot-2009, Hammerer2010,Chaneliere2018}. 
The ability of QM to synchronize operations with photonic qubits enables its application to an expanding range of problems in quantum communications and computing, such as 
 QM-enhanced quantum communication
\cite{bhaskar2020experimental}
QM-assisted multi-photon  synchronization \cite{kaneda2017quantum}, single-photon generation \cite{kaneda2019high}.
The use of QM for storing telecom-band photonic qubits encoded in various physical bases, including polarization, frequency, and time-bin states, is also expanding \cite{li2025efficient}.
Enhancing QM capabilities provides their utility for diverse tasks within the quantum internet
\cite{azuma2023quantum}.
Growing demand for higher-bandwidth quantum channels has focused attention on multi-frequency light as a crucial additional degree of freedom for quantum information transmission and processing.
Frequency-bin encoding introduces new degrees of freedom into multi-frequency light fields that will allow these fields to be used for quantum information processing, communications, and multiplexing \cite{lukens2016frequency,lu2020fully,tagliavacche2025frequency,vinet2026time}, while simultaneously posing new challenges for experimental realization \cite{lu2023frequency}.

Thus, widespread use of multimode and multifrequency light fields in optical quantum technologies requires hardware that performs fundamental photonic operations without relying on complex quantum computers. 
A more versatile QM could occupy this niche by combining data storage with elementary processing capabilities.
In particular, leveraging QM for controlled frequency-bin qubit generation would significantly advance next-generation optical quantum information processing \cite{menicucci2008one,humphreys2014continuous,lukens2016frequency,lu2023frequency} and communication systems \cite{clemmen2016ramsey,myilswamy2025chip,khodadad2025frequency,vinet2026time}.

Building upon the foundational principles of photon-echo quantum memory established in our previous works \cite{moiseev2001, Moiseev2003, Moiseev2010cavity}, the present work extends it by harnessing Pre-created Long-lived Macroscopic  coherence (PLMC) as an active control resource rather than a passive relay. 
We demonstrate that the proposed approach enables deterministic synthesis of arbitrary frequency-bin photonic qubits via robust radio-frequency (RF) rotations alone — thereby eliminating complex optical pulse shaping and transforming quantum memory into an active processor for light.
The PLMC protocol itself was proposed theoretically in our work \cite{Moiseev_PRL2025} and has recently been validated experimentally in Ref. \cite{Xu_2026_AQT}, where the authors explicitly recognize it as establishing a new paradigm for quantum memory, a conclusion independently supported in  \cite{li2026demand}.

Here we show that the developed mechanism for deterministic photonic qubit initialization relies on strong coupling between the frequency-bins via the spin PLM coherence,
imposing a generalized \textit{spectrally nonlocal} impedance-matching condition, thereby granting arbitrary control over the photonic qubit state.
We also discuss operations with frequency-bin state, as well as the experimental implementation and scalability of this approach, along with its prospective development and applications in quantum networks.

\section{Physical platform}

The proposed QM  architecture capitalizes on three core components.

1) We employ rare-earth ion (REI) doped crystals that offer unique opportunities for long-lived quantum memories \cite{Tittel-LPR-2010,Chaneliere2018,2024-Moiseev-Physics-Uspekhi}  due to their extended optical and spin coherence times \cite{GOLDNER-2015,Businger2022,Guo2023}. 
In addition, erbium ions provide operations with telecom band photons ($\lambda \approx 1550$ nm) and can be employed in cavity assisted quantum processors \cite{kinos2022high,kinos2021roadmapQC} and single-photon sources \cite{dibos2018atomic,yu2023frequency}. These properties make REI ensembles a promising platform for implementing the proposed PLMC protocol.

2) Practical implementation requires compact QM cells interoperable with integrated photonics. 
We meet this requirement using cavity-assisted storage \cite{Sabooni2013,2021-PRB-Minnegaliev,Zhou_Review_2023,Duranti2024,Kollath_2025,Liu2025,Lau2025}, based on impedance matching between the resonant mode and atomic ensemble \cite{Moiseev2010cavity,Afzelius-PRA-2010}. 
Here we derive a generalized condition enabling deterministic synthesis of arbitrary frequency-bin states.

3) In addition to the aforementioned properties, in this work we propose using spin PLM coherence and show how manipulation of this coherence allows engineering arbitrary frequency-bin photonic states.

In characterizing this third property, we address the fundamental challenge of fully exploiting quantum coherence in developing new technologies \cite{streltsov2017colloquium, chitambar2019quantum, wu2021experimental}.
The proposed use of spin PLM coherence constitutes a step toward realizing this goal. 
Crucially, in our previous work \cite{Moiseev_PRL2025}, we demonstrated that the use of spin PLM coherence endows the QM with new functionalities without introducing excess optical noise — a prediction recently confirmed experimentally in Ref. \cite{Xu_2026_AQT}.

\begin{figure}
\includegraphics[width=1.0\linewidth]{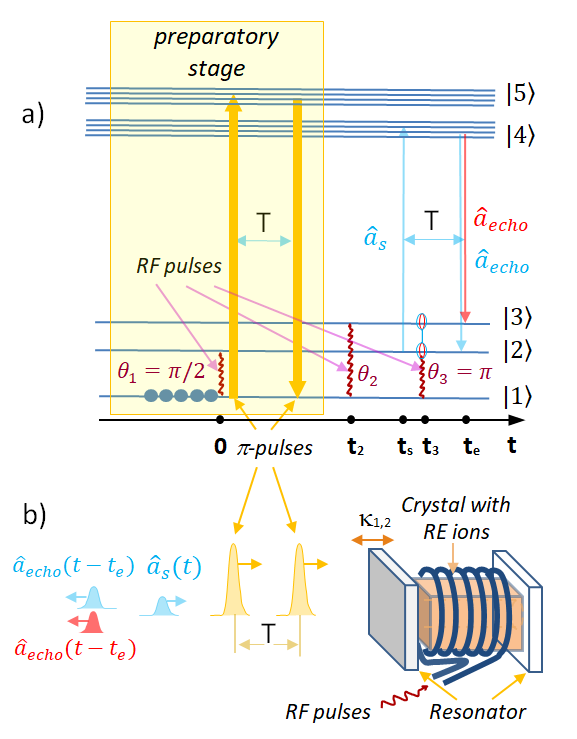}
\caption{\label{QM_scheme} 
a) Atomic energy levels of the REIs and timing sequence of control two $\pi$-laser pulses and three RF pulses with pulse areas $\theta_1$, $\theta_2$, $\theta_3$; arrival times of the incoming signal pulse $\hat{a}_s$ and the emitted frequency-bin echo $\hat{a}_{echoes}$. 
b) Simplified experimental schematic: a REIs-doped crystal is placed inside a dual-mode optical resonator; REIs are excited by two $\pi$-laser pulses and a signal pulse $\hat{a}_s$ coupled through the cavity; the emitted frequency-bin echo $\hat{a}_{echoes}$ follows; coherent spin rotation is induced by an inductively coupled RF pulses.
}
\end{figure}

The schematic of QM protocol for generating frequency-bin qubits in a five-level atomic medium is presented in Fig.\ref{QM_scheme}.
Notably, REIs with such a level structure are well-suited for quantum coherence control and are actively employed in the development of photon-echo-based QM protocols (see, e.g., recent works \cite{Guo2023,feldmann2025cavity,Xu_2026_AQT}.
The QM cell comprises a dual-mode cavity housing an REI-doped crystal.
We assume that the eigen frequencies of resonator ($\omega_2$ and $\omega_3$) tuned in resonance with optical transitions $\ket{2}\leftrightarrow\ket{4}$ and $\ket{3}\leftrightarrow\ket{4}$ ($\omega_2=\omega_{42}$ and $\omega_3=\omega_{43}$).
REI-doped crystal is placed within an inductive coil, which delivers the RF excitation field to the REI-spin transitions.
Following the approach \cite{Moiseev_PRL2025}, 
we consider REIs exhibiting hyperfine and Zeeman interactions in both the ground and excited optical states.
To simplify the analysis without loss of generality, we assume that the spin transitions of the ground level are characterized solely by homogeneous broadening in the studied ensemble of REIs. 
Moreover, this approximation is also valid for time scales comparable to and greater than the characteristic dephasing   time $T_2^*=\Delta_{in}^{-1}$ of the optical transitions depicted  in Fig. \ref{QM_scheme}, which illustrates significant inhomogeneous line broadening $\Delta_{in}$ of the optical transitions.
Under these conditions, which are consistent with experimental work \cite{ma2021elimination}, the inhomogeneous broadening of the optical transitions used is strongly correlated. 
This implies that frequency detunings across different optical transitions are nearly identical with high precision (i.e. $\Delta_{mn}^j\cong\Delta_j$ where $n=1,2,3$; $m=4,5$). 
The correlations arise because local electric field variations affect all atomic transitions equally (see below), whereas magnetic interactions have a negligible impact on the evolution of optical coherences within the relevant timescales.
A more general scenario, which takes into account and compensates for the effect of spin inhomogeneity on optical transitions, warrants separate consideration (see \cite{Moiseev_PRL2025}).

Before launching the signal pulse into the QM cell, we prepare a suitable spin PLM coherence.
Starting by initializing all atoms in the ground optical state $\ket{\Psi_a(0)}=\prod_{j=1}^N\ket{1}_j$ via standard pumping techniques, we then apply an RF pulse with a pulse area $\theta_1$ and phase $\phi_1$,
which drives the resonant spin transition $\ket{1}\leftrightarrow\ket{2}$.
Then, separated by a time delay $T$, two optical $\pi$-pulses act on the resonant optical transition $\ket{1}\leftrightarrow\ket{5}$ of REIs.
Now, complementary to PLMC protocol, the REIs are excited by the second short RF pulse at time $\approx t_2$, characterized by its pulse area $\theta_2$ and phase $\phi_2$. 
As a result, we obtain the following wavefunction for the REIs  $\ket{\tilde{\Psi}_a(t_2)}=\prod_{j=1}^N\ket{\tilde{\psi} (t_2)}_j$, where

\fla{
&\ket{\psi (t_2)}_j=-\cos\frac{\theta_2}{2} \cos\frac{\theta_1}{2} e^{-i(\omega_{51}^jT+\phi_{l,p})} \ket{1}_j
\nonumber
\\
&-i\sin\frac{\theta_1}{2}e^{i\phi_1}  \ket{2}_j+i\sin\frac{\theta_2}{2}
\cos\frac{\theta_1}{2}  
e^{-i(\omega_{51}^jT+\phi_{l,p}-\phi_2)}\ket{3}_j,
\label{psi_t_2}
}
here $\ket{\tilde{\psi}_a(t)}_j=\exp\{\frac{i}{\hbar}\hat{H}_{a,0}t\}\ket{\psi(t)}_j$, $\hat{H}_{a,0}^j=\sum_{m=1}^5E_m\ket{m}_{jj}\bra{m}$, $\ket{\tilde{\psi}_a(t)}_j$ is the wave function of the $j$-th atom after the action of the second RF pulse, $\omega_{51}^j=\omega_{51}+\Delta_j$, where $\omega_{mn}=\frac{E_m-E_n}{\hbar}$, 
$\phi_{l,p}=\phi_{l,2}-\phi_{l,1}$, where $\varphi_{l,1}$ and $\phi_{l,2}$ are the phases of the two laser pulses.

Note that, the pre-created quantum state \ref{psi_t_2} of the REIs features three spin coherences ($\hat{S}_{12}$, $\hat{S}_{13}$,$\hat{S}_{23}$) instead of a single coherence as in the original PLMC protocol \cite{Moiseev_PRL2025}.
The presence of multiple spin coherences introduces an extra control handle for the QM. 
In what follows, we show that this allows us to engineer arbitrary output states of the retrieved photonic echo.

The REIs are now primed to receive a single-mode signal pulse (photon wave packet),   into the memory cavity at $t_s$,  resonant with the $\ket{2}\leftrightarrow\ket{4}$ transition $\omega_2=\omega_{42}$ (see Fig. \ref{QM_scheme}a)).
The interaction dynamics between the signal pulse and the REIs in the resonator are described using Heisenberg equations for the cavity mode and atomic coherences within the standard input-output approach  of quantum optics \cite{Walls,gardiner2015quantum}. 
In the interaction picture, we obtain the following equations for the slowly varying atomic coherences $\hat{S}_{n4}^{j}$, $\hat{S}_{nn'}^{j}$ ($n,n'=1,2,3$, $n\neq n'$ ) and the intracavity field amplitude $\hat{a}_{2}$:

\fla{
&\frac{\partial \hat{a}_{2} }{\partial t}  = - \frac{\kappa_2}{2} \hat{a}_2  
- i \sum_{j=1}^{N} g^*_{2} \hat{S}_{24}^{j}+\sqrt{\kappa_2} \hat{a}_{in},
\label{eq_a_2}
\\
& \frac{\partial \hat{S}_{24}^{j} }{\partial t} = - (i\Delta_j+\gamma) \hat{S}_{24}^{j} 
-i g_2 \hat{a}_2 \big(\hat{S}_{22}^{j}-\hat{S}_{44}^{j}\big)+\hat{F}_{24},
\label{eq_S_24}
\\
& \frac{\partial \hat{S}_{14}^{j} }{\partial t} = - (i\Delta_j+\gamma) \hat{S}_{14}^{j} 
-i g_2 \hat{a}_2 \hat{S}_{12}^j+\hat{F}_{14},
\label{eq_S_14}
 \\
& \frac{\partial \hat{S}_{34}^{j} }{\partial t} = - (i\Delta_j+\gamma) \hat{S}_{34}^{j} 
-i g_2 \hat{a}_2 \hat{S}_{32}^j+\hat{F}_{34},
\label{eq_S_34}
 \\
& \frac{\partial \hat{S}_{nn'}^{j} }{\partial t} \cong 
- \gamma_{nn'}\hat{S}_{nn'}^j,
\label{eq_S_nn'}
}
where we neglect the effect of the weak signal pulse on both the populations of states $\ket{1}$, $\ket{2}$ and $\ket{3}$ and the spin coherence established between these states, $\gamma_{nn'}=\gamma_s$ is a weakest decay constant ($n\neq n'$), which can be neglected with fast memory operation;  $\gamma_{14}\cong\gamma_{24}\cong\gamma_{34}=\gamma$.
In Eqs.\eqref{eq_a_2}-\eqref{eq_S_nn'}, we employ the resonant interaction between the cavity mode and the REIs 
$\hat{V}_{a,f}=\hbar\sum_j\big(g_{2}\hat{a}_{2} \hat{S}_{42}^j+h.c.\big)$, where $g_{n}$ is a  coupling constant of cavity mode with REI on  $\ket{n}\leftrightarrow\ket{4}$ optical transition, $\kappa_2$ is a coupling constant of the cavity mode $\hat{a}_{2}$ with external waveguide, 
$\hat{a}_{in}(t)$ describe the operator of input signal pulse (with commutation relations $[\hat{a}_{in}(t'),\hat{a}_{in}^{\dagger}(t)]=\delta(t'-t)$),
dual mode cavity is described by the Hamiltonian  $\hat{H}_f=\omega_1\hat{a}^{\dagger}_1 \hat{a}_1+\omega_2\hat{a}^{\dagger}_2 \hat{a}_2$ (with commutation relations $[\hat{a}_{s}(t'),\hat{a}_{s'}^{\dagger}(t)]=\delta_{s,s'}$, $s,s'=1,2$).

Eqs. \eqref{eq_S_24}-\eqref{eq_S_34} show that the interaction dynamics now involve two additional optical (Raman) coherences ($\hat{S}_{34}$ and $\hat{S}_{14}$) rather than just one-$\hat{S}_{14}$ in original PLMC protocol, in addition to the primary optical coherence $\hat{S}_{24}$ of the signal-resonant transition. 
This extra Raman coherence $\hat{S}_{34}$  is excited in the ions similarly to the first one $\hat{S}_{14}$: it arises due to the resonator mode acting on the ions in the presence of an auxiliary pre-existing spin coherence $\hat{S}_{32}$. 
Mathematically, this mechanism is described by the newly derived Eq. \eqref{eq_S_34}.

Despite the increased number of equations, the resonator mode $ \hat{a}_{2}$ and the primary optical coherence $\hat{S}_{34}$ evolve independently of other coherences under strong RF driving that creates spin PLM coherence. 
This is because the additional Raman coherences $\hat{S}_{14}$ and $\hat{S}_{34}$  remain dephased during excitation; consequently, they do not establish coherent coupling with the cavity modes and thus have no effect on them.

Applying the Fourier representation to the input and cavity fields ($\hat{a}_{in,2}(t-t_s)=\int d\omega e^{-i\omega (t-t_s)}\tilde{a}_{in,2}(\omega)$),
we solve the system of Eqs. \eqref{eq_a_2}--\eqref{eq_S_24} 
for the resonator mode $\tilde{a}_{2}$ and the atomic coherences $\hat{S}_{n4}^j$ ($n=1,2,3$) in the limit $\gamma t\ll 1$, $\gamma_s t\ll 1$:

\fla{
\hat{S}_{n4}^j(t>t_s)&=-2\pi i g_2 e^{-i\Delta_j(t-t_s)} \tilde{a}_{2}(\Delta_j)\hat{S}_{n2}^j (t_s),
\label{P_n4-t}
\\
\tilde{a}_{2}(\omega)&=
\frac{\sqrt{\kappa_2} \tilde{a}_{in}(\omega)}{\frac{1}{2}\kappa_2+\varkappa_{22}(\omega,t_s)-i\omega},
\label{a_2-omega}
}
where
\fla{
\hat{S}_{12}^j(t_s)
&= i\cos(\frac{\theta_2}{2})\frac{\sin \theta_1}{2}
e^{i\varphi_{1,2}^j},
\label{P_spin_coh-12}
\\
\hat{S}_{32}^j(t_s)
&= -e^{-i\phi_2}\sin(\frac{\theta_2}{2})\frac{\sin \theta_1}{2}
e^{i\varphi_{1,2}^j},
\label{P_spin_coh-32}
\\
\hat{S}_{13}^j(t_s)
&= -ie^{i\phi_2}\frac{\sin\theta_2}{2} \cos^2(\frac{\theta_1}{2}).
\label{P_spin_coh-13}
}
We have taken into account the initial state $\ket{\Psi_a(t=t_s)}$ Eq.\eqref{psi_t_2} and $\hat{S}_{22}^j(t_s)-\hat{S}_{44}^j(t_s)\cong\hat{S}_{22}^j(t_s)=\sin^2(\frac{\theta_1}{2})$,
$\varphi_{1,2}^j=(\omega_{51}+\Delta_j)T+\phi_{l,p}+\phi_1$,
$\varkappa_{22}(\omega,t_s)=
\alpha_{22} \langle\hat{S}_{22} (t_s)\rangle$, where
$\alpha_{nn}(\omega)=\frac{N|g_{nn}|^2}{\Delta_{in}-i\omega}$.

In the derivation of Eqs. \eqref{P_n4-t}-\eqref{a_2-omega},
accounting for the macroscopic nature of the spin coherence involves replacing the spin operators with their expectation values in $\sum_{j=1}^{N} g_2 \hat{S}_{nm}^{j}$ by  $\sum_{j=1}^{N} g_2 \langle\hat{S}_{nm}^{j}\rangle$ and transferring for summation over the atomic responses to the continuous integration using Lorentzian shape of the inhomogeneous broadening of optical transition with linewidth $\Delta_{in}$.   

We focus on the case of efficient spectral storage where the spectral width $\delta\omega_s$ of the input light pulse is negligible compared to the absorption linewidth ($\delta\omega_s \ll \Delta_{in}, \kappa_{2}$).
Using the relation between the input and output fields $a_{in}+a_{out}=\sqrt{\kappa_2}a_{2}$, 
we derive the basic condition for impedance matching, which ensures efficient storage of the signal pulse into REIs: $\varkappa_{22}(0,t_0)=\alpha_{22}(0) \sin^2(\theta_1/2) =\frac{1}{2}\kappa_2$, providing $a_{out}=0$ and $a_{2}(t)=\frac{1}{\sqrt{\kappa_2}}a_{in}(t)$ (below $\alpha_{nn}(0)\equiv \alpha_{nn}$).

Taking into account the dependence of the phase $\varphi_{1,2}^j$ on the frequency detuning $\Delta_j$, we find that the optical coherences $\hat{S}_{14}^j(t>t_s)$ and $\hat{S}_{34}^j(t>t_s)$ rephase simultaneously at time $t=t_e=t_s+T$.
To ensure efficient emission of a two-color photon echo at $t = t_e$ — with carrier frequencies $\omega_2$ and $\omega_3$ coinciding with the resonator's eigenfrequencies — we apply an additional third RF $\pi$-pulse (with phase $\phi_3$) driving the $|1\rangle \leftrightarrow |2\rangle$ transition.
This pulse effects the following transformation of  spin and optical coherences:

\fla{
\hat{S}_{13}^j(\hat{S}_{14}^j)&\xrightarrow{\text{RF} (\pi,\phi_3)}-ie^{i\phi_3}\hat{S}_{23}^j(\hat{S}_{24}^j),
\nonumber
\\
\hat{S}_{23}^j(\hat{S}_{24}^j)&\xrightarrow{\text{RF} (\pi,\phi_3)}-ie^{-i\phi_3}\hat{S}_{13}^j(\hat{S}_{14}^j).
\label{swaping}
}

Eqs. \eqref{swaping} effect a swap of optical coherences $\hat{S}_{14}^j$ and $\hat{S}_{24}^j$, thereby establishing phasing coherence at the $\ket{2}\leftrightarrow\ket{4}$ transition instead of the $\ket{1}\leftrightarrow\ket{4}$ transition.
Consequently, the rephasing of atomic dipoles on the $\ket{2}\leftrightarrow\ket{4}$ and $\ket{3}\leftrightarrow\ket{4}$ transitions gives rise to a two-color photon echo ($\omega_2, \omega_3$) at $t = t_e$.
Following the derivation of Eqs. \eqref{eq_a_2}--\eqref{eq_S_34}, we obtain the following system of equations describing the emission of a photon echo:

\fla{
&\frac{\partial \hat{a}_{2,e} }{\partial t}  = - \frac{\kappa_2}{2} \hat{a}_{2,e}  
- i \sum_{j=1}^{N} g^*_{2} \hat{S}_{24}^{j},
\label{echo_a_2-echo}
\\
&\frac{\partial \hat{a}_{3,e} }{\partial t}  = - \frac{\kappa_3}{2} \hat{a}_{3,e} 
- i \sum_{j=1}^{N} g^*_{3} \hat{S}_{34}^{j},
\label{echo_a_3-echo}
\\
& \frac{\partial \hat{S}_{24}^{j} }{\partial t} = - i\Delta_j \hat{S}_{24}^{j} 
-i g_2 \hat{a}_{2,e} \hat{S}_{22}^{j}-ig_3\hat{a}_{3,e}\hat{S}_{23}^j,
\label{eq_S_24-echo}
\\
& \frac{\partial \hat{S}_{34}^{j} }{\partial t} = - i\Delta_j \hat{S}_{34}^{j} 
-i g_3 \hat{a}_{3,e} \hat{S}_{33}^j-ig_2\hat{a}_{2,e}\hat{S}_{32}^j,
\label{eq_S_34-echo}
}
where the cavity mode $\hat{a}_{3,e}$ is connected to the waveguide by a coupling rate $\kappa_3$,  $\hat{S}_{11}^j= \sin^2\frac{\theta_1}{2}$, $\hat{S}_{22}^j=\cos^2\frac{\theta_2}{2} \cos^2\frac{\theta_1}{2}$,
$\hat{S}_{33}^j=\sin^2\frac{\theta_2}{2} \cos^2\frac{\theta_1}{2}$, spin PLM coherence $S_{23}^j(t_e)=e^{i(\phi_2-\phi_3)} \frac{\sin\theta_2}{2}\cos^2\frac{\theta_1}{2}$ (see Eqs.\eqref{P_spin_coh-13}, \eqref{swaping}).

Considering the initial values of the optical coherences $S_{24}^j$ and $S_{34}^j$  for $t>t_3$ (see Eqs. \eqref{P_n4-t} \eqref{swaping}), 
we solve the linear system of Eqs. \eqref{echo_a_2-echo}-\eqref{eq_S_34-echo} in the Fourier domain to find the intracavity amplitudes $a_{2,e}(\omega)$ and $a_{3,e}(\omega)$. 
The corresponding output fields are then found via the input-output boundary conditions
$\tilde{a}_{2,e,out}=\sqrt{\kappa_2} \tilde{a}_{2,e}$ 
$\tilde{a}_{3,e,out}=\sqrt{\kappa_3} \tilde{a}_{3,e}$ (and specifically  $\tilde{a}_2=\frac{1}{\sqrt{\kappa_2}}\tilde{a}_{in}$):

\fla{
\tilde{a}_{2,e;out}(\omega)=e^{-i\phi_3}
\frac{\alpha_{22}e^{i\varphi_{1,2}}\cos\frac{\theta_2}{2}\sin\theta_1 \tilde{a}_{in}(\omega)}{\kappa_2\Bigl(\frac{1}{2}+\frac{\alpha_{22}\langle\hat{S}_{22}\rangle}{\kappa_2}+\frac{\alpha_{33}\langle\hat{S}_{33}\rangle}{\kappa_3}\Bigl)},
\label{solution-a2e=1}
\\
\tilde{a}_{3,e;out}(\omega)=e^{-i\phi_2}
\frac{\alpha_{33}(g_2/g_3)e^{i\varphi_{1,2}}\sin\frac{\theta_2}{2}\sin\theta_1 \tilde{a}_{in}(\omega)}{\sqrt{\kappa_2\kappa_3}\Bigl(\frac{1}{2}+\frac{\alpha_{22}\langle\hat{S}_{22}\rangle}{\kappa_2}+\frac{\alpha_{33}\langle\hat{S}_{33}\rangle}{\kappa_3}\Bigl)},
\label{solution-a3e=1}
}
where we again assume a sufficiently narrow input bandwidth $\delta\omega_s \ll \Delta_{in}, \kappa_{2,3}$; $\varphi_{1,2}=\omega_{51}T+\phi_{l,p}+\phi_1$ (see technical details in the End Matter).

Analyzing the found solutions \eqref{solution-a2e=1},\eqref{solution-a3e=1}, we determine that the maximum efficiency of two-color echo emission is achieved under the following conditions:

1) $\theta_1 = \pi/2$, which implies the creation of maximum spin PLM coherence and, consequently, an equal probability of finding spins in state $\ket{1}$ and $\ket{2}$ after the action of the first RF pulse. 

2) Satisfaction of a set of equalities
$\frac{\alpha_{22}}{\kappa_2}=\frac{\alpha_{33}}{\kappa_3}=1$, each of which would correspond to the fulfillment of the well-known impedance matching condition in photon-echo-based QMs \cite{Moiseev2010cavity,Afzelius-PRA-2010}, provided that all atoms are in the state $\ket{2}$ or $\ket{3}$.

Taking into account these two conditions with the population conservation law  $\langle\hat{S}_{22}\rangle+\langle\hat{S}_{22}\rangle=1/2$, 
we obtain a \textit{generalized} impedance matching condition

\fla{
\frac{\alpha_{22}\langle\hat{S}_{22}\rangle}{\kappa_2}+\frac{\alpha_{33}\langle\hat{S}_{33}\rangle}{\kappa_3}=\frac{1}{2},
\label{General_impedance_match}
}
where  $\alpha_{nn}=\frac{N|g_{nn}|^2}{\Delta_{in}}$. 

When the two aforementioned conditions leading to Eq. \eqref{General_impedance_match} are satisfied, we obtain the following expressions for the emitted echo fields:

\fla{
\tilde{a}_{2,e;out}(\omega)=e^{-i\phi_3}
e^{i\varphi_{1,2}}\cos\frac{\theta_2}{2}\tilde{a}_{in}(\omega),
\label{solution-a2e=2}
\\
\tilde{a}_{3,e;out}(\omega)=e^{-i\phi_2}
e^{i\varphi_{1,2}}\sin\frac{\theta_2}{2}\tilde{a}_{in}(\omega).
\label{solution-a3e=2}
}

Passing to the time domain for the these fields gives $\hat{a}_{echo}(t)=e^{i\phi_{\Sigma}}[\hat{a}_{e,2;out}(t;\omega_{42})+\hat{a}_{e,3;out}(t;\omega_{43})]$, ($\phi_{\Sigma}=\varphi_{1,2}-\phi_3$) where

\fla{
\hat{a}_{e,2;out}(t;\omega_{42})=&\cos(\frac{\theta_2}{2})\hat{a}_{in}(t-t_e;\omega_{42}),
\label{Two-echoes_time-1-n}
\\
\hat{a}_{e,3;out}(t;\omega_{43})=&e^{i(\phi_3-\phi_2)}
\sin(\frac{\theta_2}{2})\hat{a}_{in}(t-t_e;\omega_{43}).
\label{Two-echoes_time-2-n}
}

The appearance of two frequency components  \eqref{Two-echoes_time-1-n} and  \eqref{Two-echoes_time-2-n} in the photon echo $\hat{a}_{echo}$ demonstrates an effective method for initialization of a frequency-bin light field.
When a single-photon wave packet is used as the input signal, this protocol creates a frequency-bin photonic qubit. 
Notably, with tuned parameters of the atomic ensemble and the dual-mode resonator, the quantum state of the photonic qubit is controlled solely by setting the pulse area $\theta_2$ and phase $\phi_2$ of the RF pulse — that is, in a straightforward manner.
The ability  is based on an interesting new property of the impedance matching condition \eqref{General_impedance_match}.
Specifically, this single joint relation governs the emission of widely separated frequency components. 
This distinguishes it from conventional local impedance matching \cite{Moiseev2010cavity}, which applies independently to each spectral component of the signal field. 
Meanwhile, Eq. \eqref{General_impedance_match} reproduces previous results \cite{Moiseev2010cavity,Afzelius-PRA-2010} at $\theta_2=0,\pi$. 
This generalization stems from inter-mode coupling via  spin PLM coherence.
Owing to this interaction, Eq. \eqref{General_impedance_match} provides a \textit{spectrally nonlocal} impedance matching condition at any pulse area $\theta_2$ and $\phi_2$. 
This allows for full control over the synthesis of arbitrary frequency-bin qubits. 
In this regime, the QM operates not merely as a storage device but as an active state-processing unit, where collective spin coherence serves as a classical programmer for photonic qubits.

Before concluding, we note that the proposed QM approach can not only prepare arbitrary frequency-bin states but also store them and perform quantum processing. In the absence of $S_{23}$ coherence, this storage occurs independently for each of its spectral components when both matching conditions are met ($\frac{\alpha_{22}\langle\hat{S}_{22}\rangle}{\kappa_2}=\frac{\alpha_{33}\langle\hat{S}_{33}\rangle}{\kappa_3}=\frac{1}{2}$). 
However, when $S_{23}$ is present, efficient storage occurs under the joint condition \eqref{General_impedance_match}, which is fulfilled through the coherent interaction of the input frequency components and imprints the state onto the atomic ensemble.
By inserting a frequency-bin input into Eqs. \eqref{echo_a_2-echo}-\eqref{eq_S_34-echo}, one can find their solution — utilizing the symmetry with the zero-input solutions at the impedance-matching condition (see End Matter) — which shows that the frequency-bin field is fully absorbed by the ensemble, imprinting two optical coherences  on the $|2\rangle \leftrightarrow |4\rangle$ and $|3\rangle \leftrightarrow |4\rangle$ transitions 
with a profile matching Eqs. \eqref{P_n4-t}).
Thus, controlling $\hat{S}_{23}$ allows switching between simple and coherent storage regimes, enabling direct quantum processing of frequency-bin photonic states.
The resulting more general equations of motion, which include spin PLM coherence protocol ($S_{23}$), also exhibit a symmetry that enables reversible evolution.

\section{Discussion and Conclusion}

In summary, while the foundational principles of photon echo QM were laid down in our works \cite{moiseev2001,Moiseev2003,Moiseev2010cavity}, and extended to efficient multipulse readout in \cite{moiseev2004photon}, the current PLMC protocol establishes a new paradigm for quantum coherence control by transcending the traditional role of passive storage.
We show that harnessing Pre-created Long-lived Macroscopic (PLM) coherence fundamentally changes the very nature of light-matter interfaces. 
In the studied cavity-assisted scheme, this is realized through spectrally nonlocal impedance matching: it is the unique pre-created PLM coherence ($\hat{S}_{23}$) that enables dynamic control over the spin state via robust radio-frequency pulses — thereby mediating deterministic frequency conversion and acting as a distributed coherent quantum bus.
The identified symmetry of the equations governing the underlying quantum dynamics of QM studied can also play a crucial role in implementing other QM protocols using appropriate spin PLM coherence.
We emphasize that the generalized paradigm of QM control transforms quantum memory from a passive buffer into an active platform for in-memory photonic state processing, giving rise to previously unattainable regimes of coherent dynamics with broad technological implications.

In particular, we have shown both analytically and experimentally  \cite{moiseev2026exponential} that the new cascade modes of photon-echo formation enable exponential narrowing of the Ramsey resonance linewidth, opening up promising tools for precision metrology and atomic clocks.
Taken together with our demonstration of deterministic in-memory processing, these results establish the photon-echo framework as a versatile platform unifying quantum storage, coherent signal processing, and high-precision metrology.
Looking ahead, the active control of new PLMC states and their combination with cascade echo dynamics open a new avenue for coherent quantum processing.

Exploiting the full quantum potential of spin PLM coherence represents a particularly intriguing frontier in this research direction. 
The long-lived nature of a mesoscopic rare-earth ion spin ensemble — when it is pre-created in a squeezed spin state and strongly coupled to high-$Q$ cavity modes  — enables the deterministic preparation of complex continuous-variable cluster states of light. 
Harnessing such non-trivial many-body resources as a feed for light-matter interaction promises to transform the proposed platform into a powerful source of scalable photonic graph states — a key resource for measurement-based quantum computing. This subject will be addressed in a dedicated follow-up study.

Regarding experimental feasibility, integrated REI-doped QMs coupled to impedance-matched microcavities have already demonstrated efficiencies exceeding $80\%$ \cite{meng2026efficient}. 
Such devices have been successfully implemented within spin-wave protocols \cite{feldmann2025cavity}, where REIs are coherently mapped from excited optical states to long-lived ground-state levels. 
Furthermore, RF control fields are a well-established tool in this platform \cite{Guo2023}.
Engineering of the required high-Q cavities constitutes a solved engineering problem: promising results have already been achieved with LNOI microresonators featuring quality factors up to $1.23 \times 10^8$ at $\lambda = 1550$ nm \cite{gao2022lithium,li2023ultra}. 
These spectral parameters enable selective interaction with REI transitions, while electro-optic switches on the LNOI platform allow for precise tuning of resonator frequencies \cite{barya2025ultra,zhang2024spectral}.
Experimental conditions closely matching our proposal have already been realized in erbium-doped nanophotonic crystals \cite{Yong2023PRL} and thin-film lithium niobate platforms \cite{dutta2023atomic} 

Crucially, no fundamentally new hardware development is required: all necessary experimental tools are currently mature and readily available.

\begin{acknowledgments}
The author thanks K.I.Gerasimov and  M.M.Minnegaliev and N.S.Perminov for useful discussions.
This research was supported by the Ministry of Science and Higher Education of the Russian Federation (Reg. number NIOKTR 125012300688-6).

 \end{acknowledgments}

\onecolumngrid

\section{Appendix: Retrieval stage}

\textit{Technical details.}
The Eqs. \eqref{echo_a_2-echo}-\eqref{eq_S_34-echo} accounts for both the ground-state spin coherence $S_{23}^j(t_e)$ (see Eqs.\eqref{P_spin_coh-13}, \eqref{swaping}) and the initial  optical transitions $\hat{S}_{24}^j(t_3<t<t_e)$, $\hat{S}_{34}^j(t_3<t<t_e)$:

\fla{
\hat{S}_{24}^j(t_3<t<t_e)&=\hat{S}_{24,0}^j(t-t_e)=ie^{-i\phi_3}\cos\frac{\theta_2}{2} \pi  g_2 e^{-i\Delta_j(t-t_s)} e^{i\varphi_{1,2}^j}\sin\theta_1 \tilde{a}_{2}(\Delta_j) ,
\label{P_24-te}
}
\fla{
\hat{S}_{34}^j(t_3<t<t_e)&=\hat{S}_{34,0}^j(t-t_e)=ie^{-i\phi_2}\sin\frac{\theta_2}{2}\pi  g_2 e^{-i\Delta_j(t-t_s)} e^{i\varphi_{1,2}^j}\sin\theta_1 \tilde{a}_{2}(\Delta_j),
\label{P_34-te}
}
\fla{
S_{23}^j(t_e)&=e^{i(\phi_2-\phi_3)} \frac{\sin\theta_2}{2}\cos^2\frac{\theta_1}{2}.
\label{P_23-te}
}

Substituting Eqs. \eqref{P_23-te}, \eqref{P_34-te}  into Eqs. \eqref{eq_S_24-echo},\eqref{eq_S_34-echo}, we first find a formal solution for $\hat{S}_{24}^j(t>t_3)$, $\hat{S}_{34}^j(t>t_3)$. Inserting this result into Eqs. \eqref{echo_a_2-echo},\eqref{echo_a_3-echo} for resonator modes yields 

\fla{
&\frac{\partial \hat{a}_{2,e} }{\partial t}  = - \frac{\kappa_2}{2} \hat{a}_{2,e}  
- i \sum_{j=1}^{N} g^*_{2} \hat{S}_{24,0}^{j}(t-t_e)
-\sum_{j=1}^{N}g^*_{2}\int_{t_3}^{t}dt'e^{-i\Delta_j(t-t')}
\Bigl(g_{2}\hat{a}_{2,e}(t')\hat{S}_{22}^j(t_e)+g_{3}\hat{a}_{3,e}(t') \hat{S}_{23}^j(t_e)\Bigl),
\label{echo_a_2-echo2}
}
\fla{
&\frac{\partial \hat{a}_{3,e} }{\partial t}  = - \frac{\kappa_3}{2} \hat{a}_{3,e}  - i \sum_{j=1}^{N} g^*_{3} \hat{S}_{34,0}^{j}(t-t_e)-\sum_{j=1}^{N}g^*_{3}\int_{t_3}^{t}dt'e^{-i\Delta_j(t-t')}
\Bigl(g_{3}\hat{a}_{3,e}(t')\hat{S}_{33}^j(t_e)+g_{2}\hat{a}_{2,e}(t') \hat{S}_{32}^j(t_e)\Bigl).
\label{echo_a_3-echo2}
}

Applying the Fourier transform ($\hat{a}_{2,3;e}(t)=\int d\omega \tilde{a}_{2,3;e}(\omega)e^{-i\omega(t-t_t)}$ (and similarly to other variables) to Eqs. \eqref{echo_a_2-echo2}, \eqref{echo_a_3-echo2} under the assumption of a large Lorentzian inhomogeneous broadening of optical transitions results in

\fla{
&(\frac{\Gamma_2(\omega)}{2}-i\omega) \tilde{a}_{2,e}(\omega) +\beta_{23}(\omega) \tilde{a}_{3,e}(\omega) = - i N g^*_{2}\langle  \tilde{S}_{24,0}(\omega)\rangle,
\label{echo_a_2-echo2}
\\
&(\frac{\Gamma_3(\omega)}{2}-i\omega) \tilde{a}_{3,e}(\omega) +\beta_{32}(\omega) \tilde{a}_{2,e}(\omega) = - i N g^*_{3}\langle  \tilde{S}_{34,0}(\omega)\rangle,
\label{echo_a_3-echo2}
}

where $\frac{\Gamma_{2,3}(\omega)}{2}=\frac{\kappa_{2,3}}{2}+\frac{N|g_{2,3}|^2\langle\hat{S}_{22,33}(t_e)\rangle}{\Delta_{in}-i\omega}$,
$\beta_{32}=\frac{Ng^*_3g_2\langle\hat{S}_{32}(t_e)\rangle}{\Delta_{in}-i\omega}$,
with the following solutions:

\fla{
\tilde{a}_{2,e}(\omega)=iN\frac{\beta_{23}(\omega) g^*_{3}\langle  \tilde{S}_{34,0}(\omega)\rangle-(\frac{\Gamma_3(\omega)}{2}-i\omega)g^*_{2}\langle  \tilde{S}_{24,0}(\omega)}{\mathcal{D}(\omega)},
\\
\tilde{a}_{3,e}(\omega)=iN\frac{\beta_{32}(\omega) g^*_{2}\langle  \tilde{S}_{24,0}(\omega)\rangle-(\frac{\Gamma_2(\omega)}{2}-i\omega)g^*_{3}\langle  \tilde{S}_{34,0}(\omega)}{\mathcal{D}(\omega)},
}
where $\mathcal{D}(\omega)=(\frac{\Gamma_3(\omega)}{2}-i\omega)(\frac{\Gamma_2(\omega)}{2}-i\omega)-\beta_{23}(\omega)\beta_{32}(\omega)$.

Below we are interested for the case when 
spectral width $\delta\omega_s$ of the input light pulse is negligible compared to the absorption linewidth $\delta\omega_s\ll\Delta_{in}, \kappa_{2,3}$. 
In this case

\fla{
\tilde{a}_{2,e}(\omega)=-iN\frac{\frac{\kappa_{3}}{2}g^*_{2}\langle  \tilde{S}_{24,0}(\omega)\rangle}{\mathcal{D}(\omega)},
\\
\tilde{a}_{3,e}(\omega)=-iN
\frac{\frac{\kappa_{2}}{2}g^*_{3}\langle  \tilde{S}_{34,0}(\omega)\rangle}{\mathcal{D}(\omega)},
}

where
\fla{
\langle\tilde{S}_{24,0}(\omega)\rangle&=i\frac{g_2}{\Delta_{in}}
e^{-i\phi_3}\cos\frac{\theta_2}{2} e^{i\varphi_{1,2}}\sin\theta_1 \tilde{a}_{2}(\omega),
\\
\langle\tilde{S}_{34,0}(\omega)\rangle&=i\frac{g_2}{\Delta_{in}}
e^{-i\phi_2}\sin\frac{\theta_2}{2} e^{i\varphi_{1,2}}\sin\theta_1 \tilde{a}_{2}(\omega),
}

\fla{
\mathcal{D}(0)=\frac{\Gamma_2(0)\Gamma_3(0)}{4} -|\beta_{23}|^2
=\frac{\kappa_2\kappa_3}{2}\Bigl(\frac{1}{2}+\frac{\alpha_{22}\langle\hat{S}_{22}(t_e)\rangle}{\kappa_2}+\frac{\alpha_{33}\langle\hat{S}_{33}(t_e)\rangle}{\kappa_3}\Bigl),
}
where $\alpha_{nn}=\frac{N|g_n|^2}{\Delta_{in}}$, $\varphi_{1,2}=\omega_{51}T+\phi_{l,p}+\phi_1$.

Then after algebraic calculations, we find:

\fla{
\tilde{a}_{2,e}(\omega)=e^{-i\phi_3}\cos\frac{\theta_2}{2}
\frac{\alpha_{22}e^{i\varphi_{1,2}}\sin\theta_1 \tilde{a}_{2}(\omega)}{\kappa_2\Bigl(\frac{1}{2}+\frac{\alpha_{22}\langle\hat{S}_{22}(t_e)\rangle}{\kappa_2}+\frac{\alpha_{33}\langle\hat{S}_{33}(t_e)\rangle}{\kappa_3}\Bigl)},
\\
\tilde{a}_{3,e}(\omega)=e^{-i\phi_2}\sin\frac{\theta_2}{2}
\frac{\alpha_{33}(g_2/g_3)e^{i\varphi_{1,2}}\sin\theta_1 \tilde{a}_{2}(\omega)}{\kappa_3\Bigl(\frac{1}{2}+\frac{\alpha_{22}\langle\hat{S}_{22}(t_e)\rangle}{\kappa_2}+\frac{\alpha_{33}\langle\hat{S}_{33}(t_e)\rangle}{\kappa_3}\Bigl)}.
}

Now by applying the input-output relations 
($\tilde{a}_{2,e,out}=\sqrt{\kappa_2} \tilde{a}_{2,e}$ 
$\tilde{a}_{3,e,out}=\sqrt{\kappa_3} \tilde{a}_{3,e}$ 
with $\tilde{a}_{in}=0$ and  $\tilde{a}_2=\frac{1}{\sqrt{\kappa_2}}\tilde{a}_{in}$), 
we find the echo fields in the waveguide  (see  Eqs. \eqref{solution-a2e=1}, \eqref{solution-a3e=1}):

\fla{
\tilde{a}_{2,e;out}(\omega)=e^{-i\phi_3}\cos\frac{\theta_2}{2}
\frac{\alpha_{22}e^{i\varphi_{1,2}}\sin\theta_1 \tilde{a}_{in}(\omega)}{\kappa_2\Bigl(\frac{1}{2}+\frac{\alpha_{22}\langle\hat{S}_{22}(t_e)\rangle}{\kappa_2}+\frac{\alpha_{33}\langle\hat{S}_{33}(t_e)\rangle}{\kappa_3}\Bigl)},
\\
\tilde{a}_{3,e;out}(\omega)=e^{-i\phi_2}\sin\frac{\theta_2}{2}
\frac{\alpha_{33}(g_2/g_3)e^{i\varphi_{1,2}}\sin\theta_1 \tilde{a}_{in}(\omega)}{\sqrt{\kappa_2\kappa_3}\Bigl(\frac{1}{2}+\frac{\alpha_{22}\langle\hat{S}_{22}(t_e)\rangle}{\kappa_2}+\frac{\alpha_{33}\langle\hat{S}_{33}(t_e)\rangle}{\kappa_3}\Bigl)}.
}

\textit{Storage of frequency-bin field} - 
By assuming that the frequency-bin field  is launched into the QM cell, we insert $\hat{a}_{e,2;out}(t)$ and $\hat{a}_{e,3;out}(t)$ in Eqs. \eqref{echo_a_2-echo}, \eqref{echo_a_3-echo} we get the following equations for the cavity modes $\hat{a}_{2}$ and $\hat{a}_{3}$:

\fla{
\frac{\partial \hat{a}_{2} }{\partial t}  = - \frac{\kappa_2}{2} \hat{a}_{2}  
- i \sum_{j=1}^{N} g^*_{2} \hat{S}_{24}^{j}+\sqrt{\kappa_2} \hat{a}_{e,2;out}(t),
\label{fr-bin_a_2}
}
\fla{
\frac{\partial \hat{a}_{3} }{\partial t}  = - \frac{\kappa_3}{2} \hat{a}_{3} 
- i \sum_{j=1}^{N} g^*_{3} \hat{S}_{34}^{j}+\sqrt{\kappa_3} \hat{a}_{e,3;out}(t).
\label{fr-bin_a_3}
}

Taking into account the impedance matching condition, we impose the condition of complete absorption \eqref{General_impedance_match} for the frequency bin field (see Eqs. \eqref{Two-echoes_time-1-n}, \eqref{Two-echoes_time-2-n}),  and as a result, we obtain the following system of equations:

\fla{
&\frac{\partial \hat{a}_{2} }{\partial t}  =  \frac{\kappa_2}{2} \hat{a}_{2}  
- i \sum_{j=1}^{N} g^*_{2} \hat{S}_{24}^{j},
\label{fr-bin_a_2},
\\
&\frac{\partial \hat{a}_{3} }{\partial t}  = \frac{\kappa_3}{2} \hat{a}_{3} 
- i \sum_{j=1}^{N} g^*_{3} \hat{S}_{34}^{j},
\label{fr-bin_a_3}
}
\fla{
& \frac{\partial \hat{S}_{24}^{j} }{\partial t} = - i\Delta_j \hat{S}_{24}^{j} 
-i g_2 \hat{a}_{2} \hat{S}_{22}^{j}-ig_3\hat{a}_{3}\hat{S}_{23}^j,
\label{fb_S_24}
}
\fla{
\frac{\partial \hat{S}_{34}^{j} }{\partial t} = - i\Delta_j \hat{S}_{34}^{j} 
-i g_3 \hat{a}_{3} \hat{S}_{33}^j-ig_2\hat{a}_{2}\hat{S}_{32}^j,
\label{fb_S_34}
}
where the optical coherences initially vanish ($\langle\hat{S}_{24}^{j}(t\rightarrow -\infty) \rangle=\langle\hat{S}_{34}^{j}(t\rightarrow -\infty) \rangle=0$).

Under time reversal ($t \to -t$) and the substitution $\{\hat{S}_{24}^{j}(\Delta_j),\hat{S}_{34}^{j}(\Delta_j)\}\to \{-\hat{S}_{24}^{j}(-\Delta_j), -\hat{S}_{34}^{j}(-\Delta_j)\}$, Eqs. \eqref{fr-bin_a_2}-\eqref{fb_S_34} governing frequency-bin absorption fully coincide with Eqs. \eqref{echo_a_2-echo}-\eqref{eq_S_34-echo} describing their emission.
This implies that the equations governing the studied PLMC protocol possess a specific symmetry, thereby enabling reversible quantum evolution. 
This symmetry generalizes previous results for the CRIB protocol \cite{Kraus2006,ESMoiseev_2013} to the case with non-zero spin coherence $\hat{S}_{23}$.

\twocolumngrid

\bibliography{apssamp.bib}

\end{document}